\documentstyle[sprocl,psfig]{article}

\begin{document}
\title{TOWARDS NUMERICAL AND ANALYTICAL STUDIES OF
FIRST ORDER PHASE TRANSITIONS \footnote{To appear in the Proceedings
of the 18th Texas Symposium on Relativistic Astrophysics, eds.\ A.\ Olinto,
J.\ Frieman, and D.\ Schramm, World Scientific, Singapore.}}
\author{ Hans-Reinhard M\"ULLER }
\address{Department of Physics and Astronomy, Dartmouth College,\\
Hanover, NH 03755-3528, USA\\
{\rm (DART-HEP-97/03~~hep-lat/9704003)}}
\maketitle\abstracts{
Discrete lattice simulations of a one-dimensional $\phi^4$ theory coupled to an
external heat bath are being carried out.
Great care is taken to remove the effects of lattice
discreteness and finite size and to establish the correct correspondence
between simulations and the desired,
finite-temperature continuum limit.}

\section{Introduction}

To be able to study numerically certain properties of cosmological
first order phase transitions
such as the nucleation rate of bubbles,
we investigate first
the effects that finiteness and discreteness of a lattice
have on results derived from simulations.
For this basic study we limit ourselves to one spatial dimension
and to a real scalar field $\phi (x,t)$
subjected to a potential $V_0 (\phi)$
and an environmental temperature $T$.
The dynamics obey a Langevin equation
as given in Borrill and Gleiser~\cite{2Dmatch}.
This 2D-paper indicates
that for a given tree-level potential,
results obtained numerically on the lattice 
can't straightforwardly be identified
with analytical results.
They employ a renormalization procedure
to get rid of the lattice spacing dependence and
to identify the correct continuum limit of the simulations.
Similar problems in 1D are treated
in what follows.

\section{Method}

For classical
field theories, the one-loop corrected effective potential
is given by a momentum integral ~\cite{CFT}, and evaluated to
\begin{equation}
V_{\rm 1L}(\phi)=V_0(\phi) + {T\over 2}\int_0^{\infty}{{dk}\over
{2\pi}}{\rm ln}\left [1 + {V_0''(\phi)\over k^2} \right ] =
V_0(\phi) + {T\over 4}\sqrt{V_0''(\phi)} ~.
\end{equation}

The discretization $\delta x$ of the lattice
and its finite size $L$
introduce short and long momentum cutoffs
$k_{\rm min} = 2 \pi / L$ and 
$\Lambda = \pi / \delta x$.
Therefore the simulation only sees $\tilde V_{\rm 1L} =V_0(\phi) +
(T/2)\int_{k_{\rm min}}^{\Lambda}\ldots dk$.
If one neglects the effect of $k_{\rm min}$ which is possible
for sufficiently large $L$~\footnote{Lattice simulations only know one
size parameter, the number of degrees of freedom $N = L/\delta x$.
With a given $N$ it is always possible to choose $L$ big enough
for the effects of $\delta x$ to dominate over those of $L$.},
integration and expansion in powers of $V_0''/\Lambda^2$ (possible
for sufficiently large $\Lambda$) yields
\begin{equation}
\tilde V_{\rm 1L}=V_0 + {T\over 4}\sqrt{V_0''} -
{T\over 4\pi}  {V_0''\over \Lambda} +
\Lambda\,\, O\left( V_0''^2\over \Lambda^4 \right) ~.
\end{equation}
As is to be expected for a one dimensional system,
the limit $\Lambda \rightarrow \infty$ exists and is well-behaved;
there is no need for renormalization due to divergences.
However, the effective one-loop potential
is lattice-spacing dependent through the explicit appearance
of $\Lambda$, and so are the corresponding numerical
simulations, as evidenced in the tree-level
potential cases of Fig.\ 1 (left graphs).

Similar to the renormalization procedure for 2D systems given by
Borrill and Gleiser~\cite{2Dmatch} we remove
this dependence on $\delta x$
by adding counterterms to the tree-level potential $V_0$.
In contrast to higher-dimensional systems, these
counterterms are {\em finite},
namely $V_{\rm CT}(\phi)=
(T/ 4\pi)  (V_0''(\phi)/ \Lambda)$.
Hence the lattice simulation works with the corrected potential
\begin{equation}\label{e.latt}
V_{\rm Latt}(\phi)=V_0(\phi) +
{T V_0''(\phi)\over 4\pi^2}  \delta x 
\end{equation}
and simulates the continuum limit to one loop,
$\tilde V_{\rm 1L}(\phi) = V_0 + (T/2) \int_{0}^{\Lambda}\ldots dk =
V_{\rm 1L}(\phi)$ (where $V_{\rm Latt}''$ is employed in the integrand),
just as it should be.

\section{Application}

Since the numerical extraction of bubble nucleation rates
is a contrived process the ideas of Section 2 are tested initially
with the {\it symmetric} double well potential
$V_0(\phi) = (\lambda / 4) \left( \phi^2 - \phi_0^2 \right) ^2$.
We compare simulations using $V_0$ alone with those
employing $V_{\rm Latt}(\phi) =
V_0(\phi) + 3\lambda T \delta x \phi^2/4 \pi^2$ (eq.\ \ref{e.latt}).
One set of runs investigates the mean field value
$\langle \bar \phi \rangle$ of the metastable
equilibrium before the first kink-antikink pair occurs
($\bar\phi (t) = (1/L) \int\phi (x,t) dx$).
Another set of runs measures the kink-antikink pair density $n_{\rm p}$
(proportional to the number of zeros of the low-pass filtered field).
Fig.\ 1 shows the comparison for different lattice spacings $\delta x$.
Apart from a discrepancy for very coarse grids 
($\delta x \approx 1$) the average field
value is clearly lattice-spacing independent in the right panel
($V_{\rm Latt}$),
in contrast to the use of $V_0$.
The effect is even more striking in the case of $n_{\rm p}$
where the addition of the finite counterterm 
removes any $\delta x$ dependence.

In summary it was demonstrated that
even in a field theory without divergences finite counterterms
play a role. Their inclusion gets rid of the
dependence of simulations on size and lattice spacing.
This can be observed in the averaged field value
and in the density of kink-antikinks.
Further studies of this renormalization procedure \cite{kink}
identify the correct continuum
limit of simulations, thus matching theory and numerical results.

\section*{Acknowledgments} The
author thanks Marcelo Gleiser for suggesting the topic
and valuable discussions.
Support from
NASA  grant NAGW-4270 and NSF grant
PHY-9453431 is gratefully acknowledged.

\begin{figure} 
\begin{minipage}[b]{.49\linewidth}
\psfig{figure=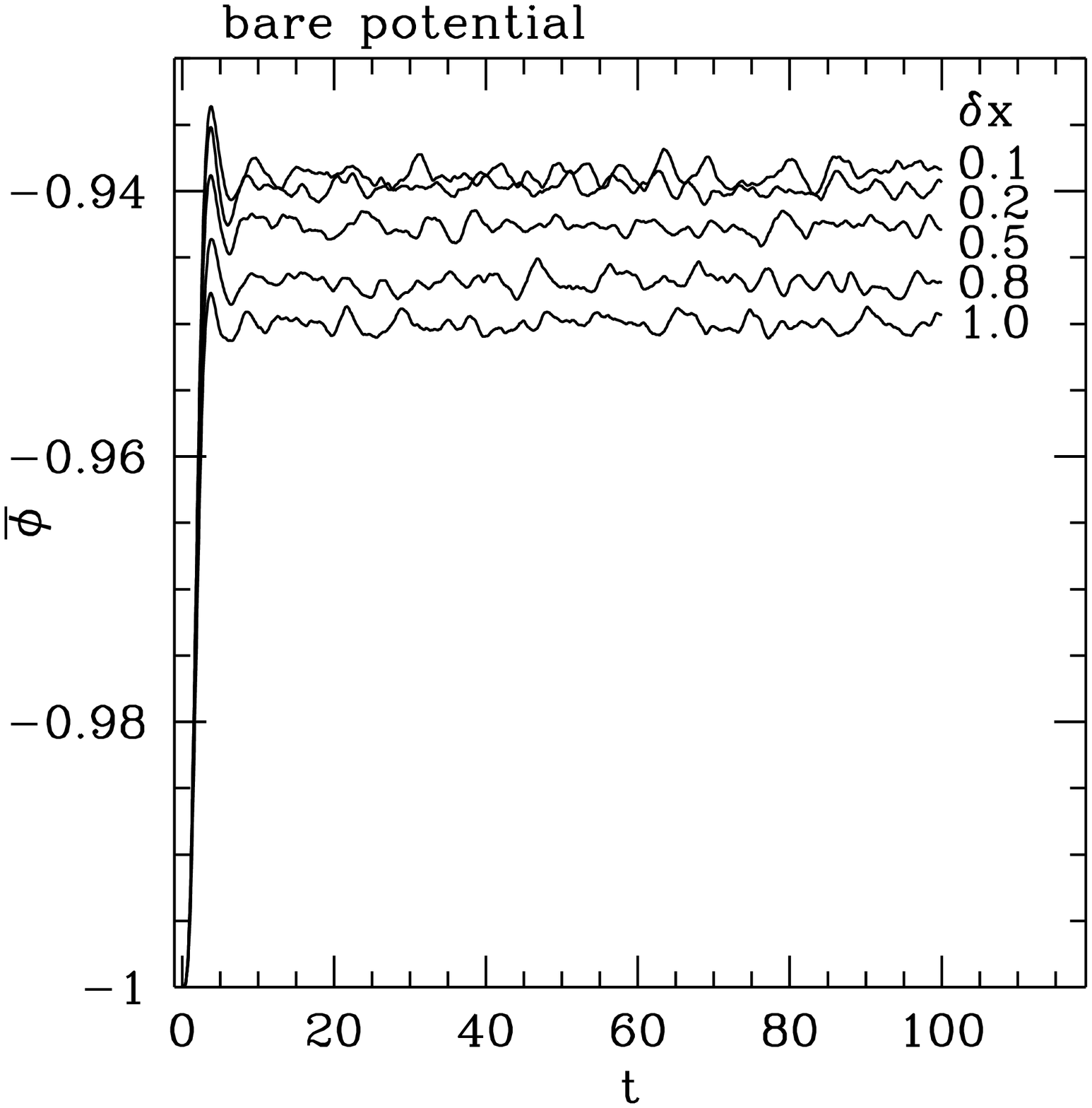,width=\linewidth,height=1.8in}
\end{minipage}
\begin{minipage}[b]{.49\linewidth}
\psfig{figure=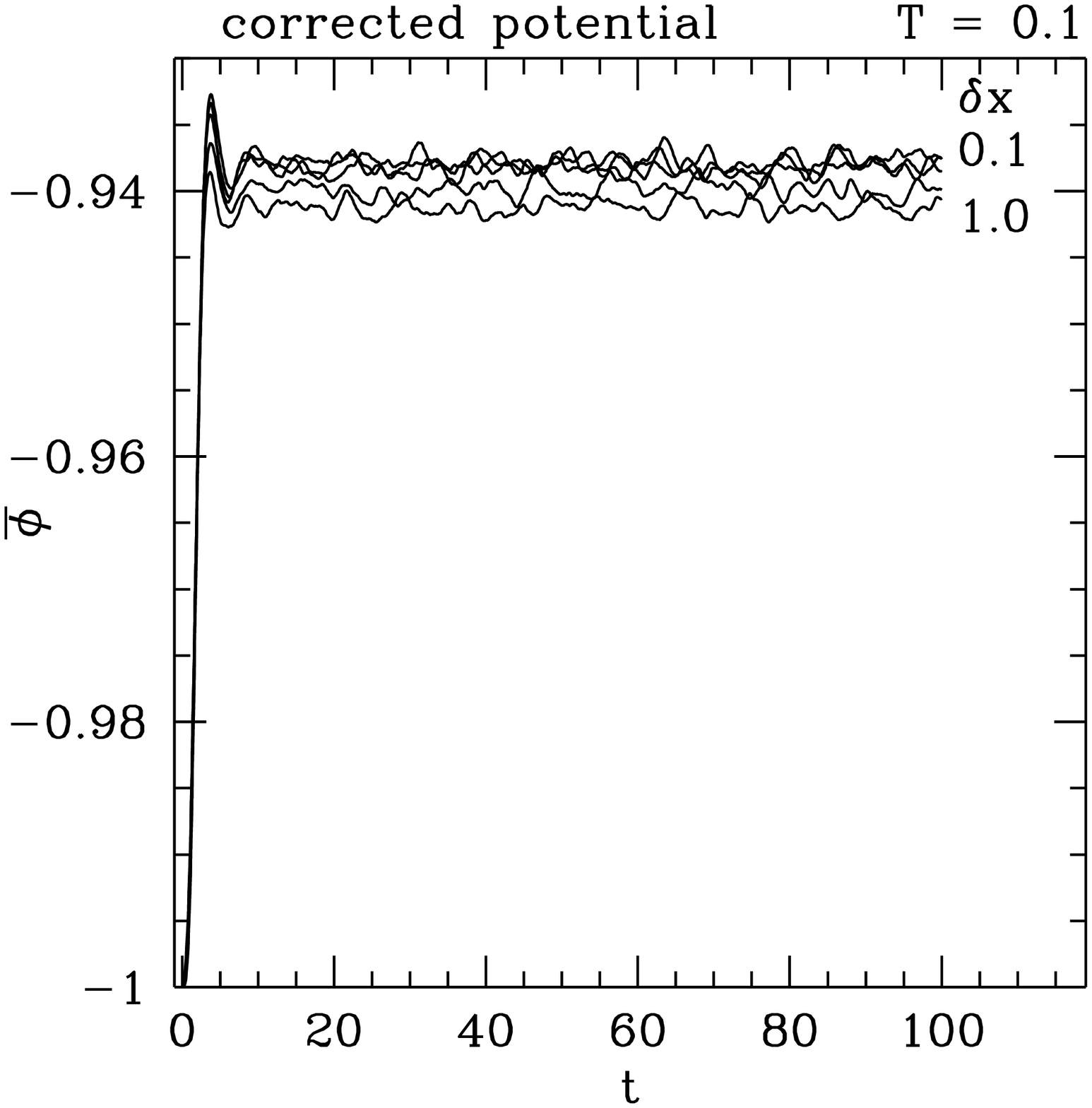,width=\linewidth,height=1.8in}
\end{minipage}
\begin{minipage}[b]{.49\linewidth}
\psfig{figure=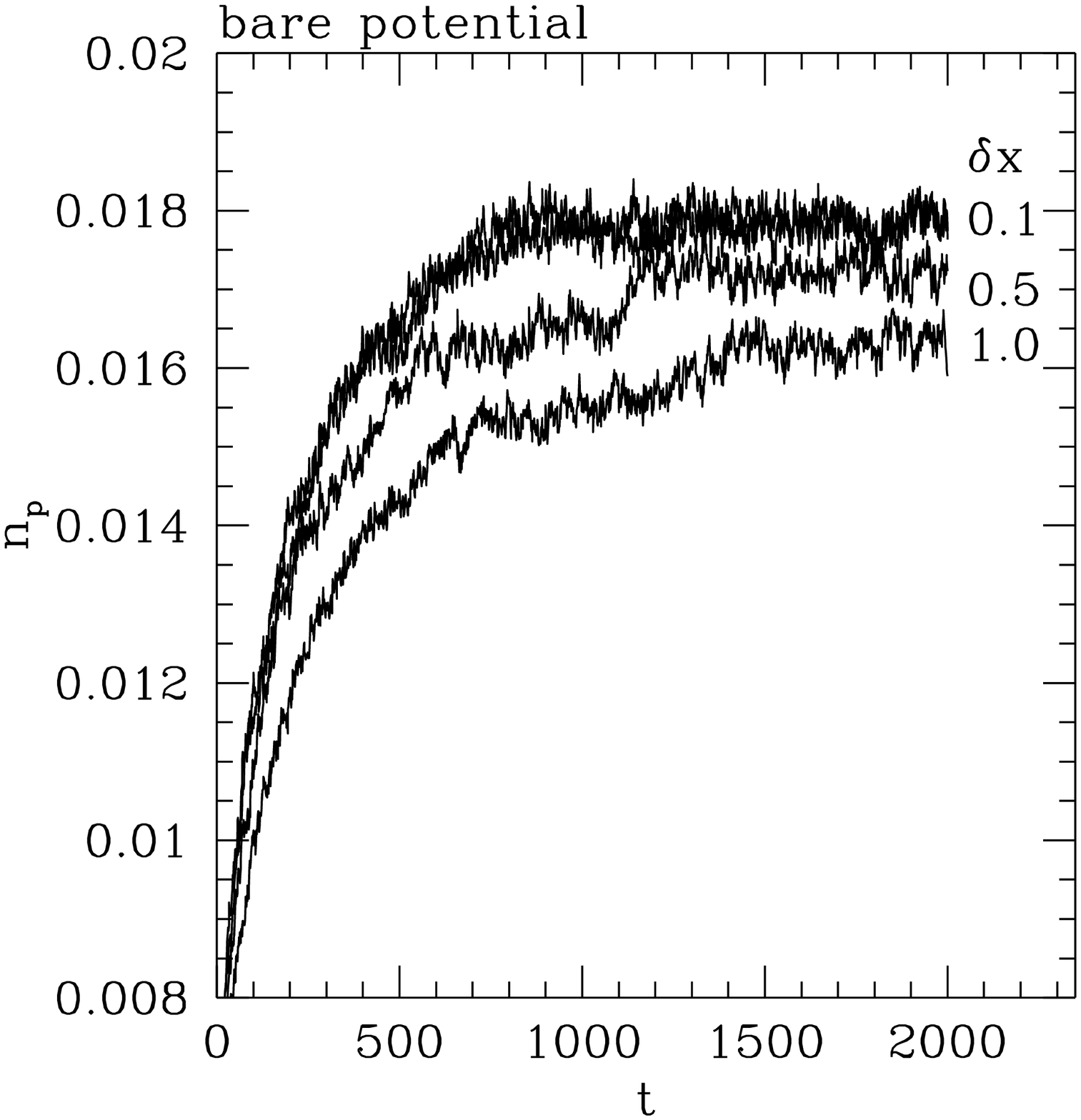,width=\linewidth,height=1.8in}
\end{minipage}
\begin{minipage}[b]{.49\linewidth}
\psfig{figure=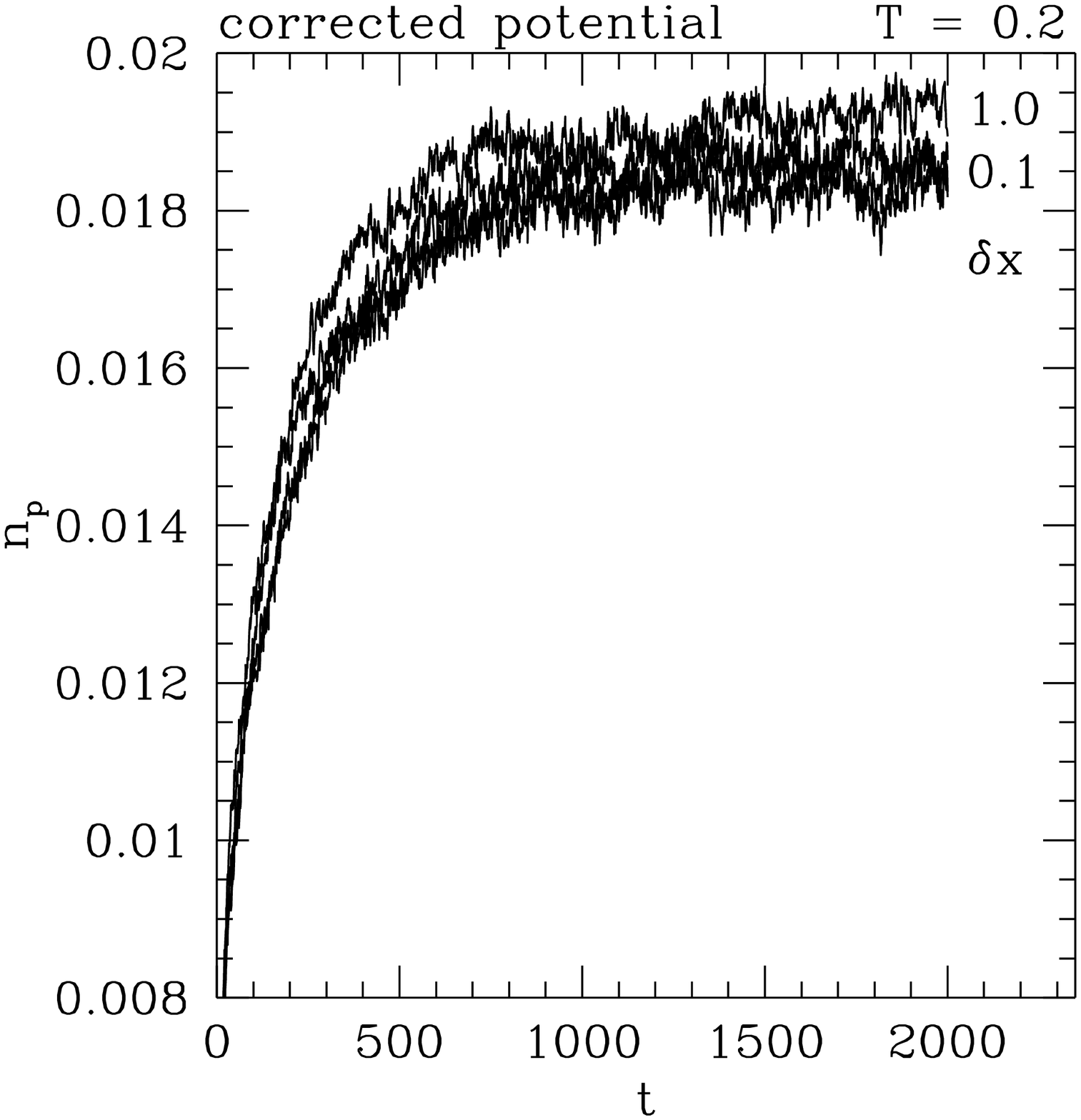,width=\linewidth,height=1.8in}
\end{minipage}
\caption[]{Average field value $\bar\phi (t)$, top,
and density of kink-antikinks (half of density of zeros), bottom,
using the tree-level potential,
left, and the corrected potential, right.
\label{fig:pairs}}
\end{figure}

\end{document}